RevTex

\documentstyle[floats,prl,aps]{revtex}
\begin{document}
\renewcommand{\thefootnote}{\fnsymbol{footnote}}
\draft
\title{\large\bf 
Integrable Kondo impurities in the one-dimensional supersymmetric
extended Hubbard model} 

\author { Huan-Qiang Zhou
\footnote {E-mail: hqz@maths.uq.edu.au},
Xiang-Yu Ge 
\footnote {E-mail: xg@maths.uq.edu.au},
and Mark D. Gould } 

\address{      Department of Mathematics,University of Queensland,
		     Brisbane, Qld 4072, Australia}

\maketitle

\vspace{10pt}

\begin{abstract}
An integrable Kondo problem in the one-dimensional supersymmetric extended
Hubbard model is studied by means of the boundary graded quantum inverse
scattering method.
The boundary $K$ matrices depending on the local moments of the impurities 
are presented  as a nontrivial realization of the graded reflection equation
algebras in a two-dimensional impurity Hilbert space.
Further,the model is solved by using the algebraic Bethe ansatz method
and the Bethe ansatz equations are obtained. 
\end{abstract}




\def\a{\alpha}
\def\b{\beta}
\def\d{\delta}
\def\e{\epsilon}
\def\g{\gamma}
\def\k{\kappa}
\def\l{\lambda}
\def\o{\omega}
\def\t{\theta}
\def\s{\sigma}
\def\D{\Delta}
\def\L{\Lambda}


\def\beq{\begin{equation}}
\def\eeq{\end{equation}}
\def\bea{\begin{eqnarray}}
\def\eea{\end{eqnarray}}
\def\ba{\begin{array}}
\def\ea{\end{array}}
\def\no{\nonumber}
\def\le{\langle}
\def\re{\rangle}
\def\lt{\left}
\def\rt{\right}

\newcommand{\reff}[1]{eq.~(\ref{#1})}

\vskip.3in
Kondo problem describing the effect due to the exchange interaction
between the magnetic impurity and the conduction electrons plays a very
important role in condensed matter physics \cite {K64}. Wilson \cite
{W75} developed a very powerful numerical renormalization group approach
,and the model was also solved by the coordinate Bethe ansatz method
\cite {W83,A83} which gives the specific heat and magnetization. More
recently,a conformal field theory approach was developed by Affleck and
Ludwig \cite {AL} based on a previous work by Nozi{\`e}res \cite {N74}. In the
conventional Kondo problem, the interaction between the conduction
electrons is dicarded, due to the fact that the interacting electron
system can be described by Fermi liquid.
Recently,much attention has been paid to the study
of the theory of magnetic impurities in Luttinger liquids 
(see eg. refs\cite {LT92,FN94,FJ95}).
Although some powerful methods,such as the bosonization method,boundary
conformal field theory,and the density
matrix renormalization group method,are available to help us  gain an
understanding of the critical behaviour
of Kondo impurities coupled to a Fermi or Luttinger liquid,some simple
integrable models which 
allow exact solutions are still  desirable.

Several integrable magnetic or nonmagnetic impurity problems describing
a few impurities embedded
in some correlated electron systems have so far appeared in the
literature. Among them are several versions of the supersymmetric $t$-$J$
model with impurities \cite {BAR94,BEF97,SZ97,LF98}. Such an idea to incorporate an impurity into a
closed chain may date back to Andrei and Johannesson \cite {AJ84}(see
also \cite {LS88,ZJ89}). However,the model
thus constructed suffers the lack of backward scattering and results in
a very complicated Hamiltonian which is difficult to be justified on the
physical ground. Therefore, as observed by Kane and
Fisher \cite {KF92},it seems to be advantageous to adopt open boundary 
conditions with
the impurities situated at the ends of the chain when studying
Kondo impurities coupled to
integrable strongly correlated electron systems 
\cite {Z97,PW97,ZM98}. 

In this Letter, an
integrable Kondo problem in  the 1D supersymmetric extended Hubbard
model is studied . 
It should be emphasized that the new non-c-number boundary 
K matrices arising from our approach are highly nontrivial, 
in the sense that they can not be factorized into the product of a
c-number boundary K matrix and the corresponding local monodromy
matrices. The model is solved by means of the 
algebraic Bethe ansatz method and
the Bethe ansatz equations are derived.

Let $c_{j,\s}$ and $c_{j,\s}^{\dagger}$ denote electronic creation and
annihilation operators for spin $\s$ at
site $j$, which satisfy the anti-commutation relations 
$\{c_{i,\s}^\dagger, c_{j,\tau}\}=\d_{ij}\d_{\s\tau}$, where 
$i,j=1,2,\cdots,L$ and $\s,\tau=\uparrow,\;\downarrow$. We consider 
the following Hamiltonian which describes two impurities coupled to 
the supersymmetric extended Hubbard open chain ,
\bea
H&=&-\sum _{j=1,\s}^{L-1}
(c_{j,\s}^\dagger c_{j+1,\s}+{\rm H.c.})
  (1-n_{j,-\s}-n_{j+1,-\s})
  \no\\
   & &-\sum ^{L-1}_{j=1}(c_{j,\uparrow}^\dagger c_{j,\downarrow}^\dagger
   c_{j+1,\downarrow}c_{j+1,\uparrow}
   +{\rm H.c}) 
   +2\sum ^{L-1}_{j=1}({\bf S}_j\cdot {\bf S}_{j+1}
 -\frac{1}{4}n_jn_{j+1})\no\\
 & & +J_a {\bf S}_1 \cdot {\bf S}_a +V_a n_1 +U_a n_{1\uparrow}
 n_{1\downarrow}
  +J_b {\bf S}_L \cdot {\bf S}_b +V_b n_L +U_b n_{L\uparrow}
 n_{L\downarrow},
  ,\label{ham}
\eea
where $J_g,V_g $ and $U_g (g=a,b)$ 
are the Kondo coupling constants ,the
impurity scalar potentials and the boundary Hubbard-like interaction
constants,respectively; ${\bf S}_j=\frac {1}{2}\sum
_{\s,\s'}c^\dagger_{j\s}{\bf \s}_{\s\s'}c_{i\s'}$ is the spin operator
of the conduction electrons; ${\bf S}_{g} (g = a,b)$ are the local
moments with spin-$\frac {1}{2}$ located at the left and right ends of
the system respectively;
 $n_{j\s}$ is the number density operator
$n_{j\s}=c_{j\s}^{\dagger}c_{j\s}$,
$n_j=n_{j\uparrow}+n_{j\downarrow}$.

The supersymmetry algebra underlying the bulk
Hamiltonian of this model is
$gl(2|2)$,and the integrability of the model on a closed chain has been
extensively studied by Essler , Korepin and Schoutens \cite {EKS92}
. It is quite interesting to note that although
the introduction  of the impurities spoils the supersymmetry,there is
still a remaining  $u (2)\otimes u(2)$ symmetry in the Hamiltonian (\ref {ham}).
As a result,one may add some terms like the Hubbard interaction 
$U \sum _{j=1}^L n_{j\uparrow}n_{j\downarrow}$,the chemical potential
term $\mu \sum ^L_{j=1}n_j$ and the external magnetic field $h \sum _{j=1}^L
(n_{j\uparrow}-n_{j\downarrow})$ to the Hamiltonian (\ref {ham}),without
spoiling the integrability. This explains why the model is so named
(also called the EKS model). Below we will establish the quantum integrability of the Hamiltonian
(\ref{ham}) for a special choice of the model parameters $J_g$,
$V_g$,and $U_g$
\beq
J_g = -\frac {2}{c_g(c_g+2)},
V_g = -\frac {2c_g^2+2c_g-1}{2c_g(c_g+2)},
U_g = -\frac {1-c^2_g}{c_g(c_g+2)},
\eeq
This is achieved by showing that it can be derived from
the (graded) boundary quantum inverse scattering method \cite
{Zhou97,BRA98}. 

Let us recall that the Hamiltonian of
the 1D  supersymmetric extended Hubbard  model with the periodic
boundary conditions
commutes with the transfer matrix, which is the supertrace of the
monodromy matrix $ T(u) = R_{0L}(u)\cdots R_{01}(u)$. 
Here the quantum R-matrix 
$ R_{0j}(u)$ takes the form, 
\beq
R=\frac {u-2P}{u-2}, \label {r}
\eeq
where $u$ is the spectral parameter,and $P$ denotes the graded
permutation operator,
and the subscript $0$ denotes the 4-D auxiliary superspace $V=C^{2,2}$ with
the grading $[i]=0~ {\rm if} ~i=1,2$ and $1~ {\rm if}~ i=3,4$.
It should be noted that the supertrace
is carried out for the auxiliary superspace $V$.
The elements of the supermatrix $T(u)$ are the generators
of an associative superalgebra ${\cal A}$ defined by the relations
\beq
R_{12}(u_1-u_2) \stackrel {1}{T}(u_1) \stackrel {2}{T}(u_2) =
   \stackrel {2}{T}(u_2) \stackrel {1}{T}(u_1)R_{12}(u_1-u_2),\label{rtt-ttr} 
\eeq
where $\stackrel {1}{X} \equiv  X \otimes 1,~
\stackrel {2}{X} \equiv  1 \otimes X$
for any supermatrix $ X \in End(V) $. For later use, we list some useful
properties enjoyed by the R-matrix:
(i) Unitarity:   $  R_{12}(u)R_{21}(-u) = \rho (u)$ and (ii)
 Crossing-unitarity:  $  R^{st_2}_{12}(-u)R^{st_2}_{21}(u) =
         \tilde {\rho }(u)$
with $\rho (u),\tilde \rho (u)$ being some  scalar functions.

In order to describe integrable electronic models on  open
chains, we introduce two associative
superalgebras ${\cal T}_-$  and ${\cal T}_+$ defined by the R-matrix
$R(u_1-u_2)$ and the relations \cite {Zhou97,BRA98} 
\beq
R_{12}(u_1-u_2)\stackrel {1}{\cal T}_-(u_1) R_{21}(u_1+u_2)
  \stackrel {2}{\cal T}_-(u_2)
=  \stackrel {2}{\cal T}_-(u_2) R_{12}(u_1+u_2)
  \stackrel {1}{\cal T}_-(u_1) R_{21}(u_1-u_2),  \label{reflection1}
\eeq
\bea
&&R_{21}^{st_1 ist_2}(-u_1+u_2)\stackrel {1}{\cal T}_+^{st_1}
  (u_1) R_{12}(-u_1-u_2)
  \stackrel {2}{\cal T}_+^{ist_2}(u_2)\no\\
&&~~~~~~~~~~~~~~~=\stackrel {2}{\cal T}_+^{ist_2}(u_2) R_{21}(-u_1-u_2)
  \stackrel {1}{\cal T}_+^{st_1}(u_1) R_{12}^{st_1 ist_2}(-u_1+u_2),
  \label{reflection2}
\eea
respectively. Here the supertransposition $st_{\alpha}~(\alpha =1,2)$ 
is only carried out in the
$\alpha$-th factor superspace of $V \otimes V$, whereas $ist_{\alpha}$ denotes
the inverse operation of  $st_{\alpha}$. By modifying Sklyanin's 
arguments \cite{Skl88}, one
may show that the quantities $\tau(u)$ given by
$\tau (u) = str ({\cal T}_+(u){\cal T}_-(u))$
constitute a commutative family, i.e.,
        $[\tau (u_1),\tau (u_2)] = 0$. 

One can obtain a class of realizations of the superalgebras ${\cal T}_+$  and
${\cal T}_-$  by choosing  ${\cal T}_{\pm}(u)$ to be the form
\beq
{\cal T}_-(u) = T_-(u) \tilde {\cal T}_-(u) T^{-1}_-(-u),~~~~~~ 
{\cal T}^{st}_+(u) = T^{st}_+(u) \tilde {\cal T}^{st}_+(u) 
  \lt(T^{-1}_+(-u)\rt)^{st}\label{t-,t+} 
\eeq
with
\beq
T_-(u) = R_{0M}(u) \cdots R_{01}(u),~~~~
T_+(u) = R_{0L}(u) \cdots R_{0,M+1}(u),~~~~ 
\tilde {\cal T}_{\pm}(u) = K_{\pm}(u),
\eeq
where $K_{\pm}(u)$, called boundary K-matrices, 
are representations of  ${\cal T}_{\pm}  $ in some representation
superspace. Although many attempts have been made to find c-number
boundary K matrices,which may be referred to as the fundamental
representation,it is no doubt very interesting to search for
non-c-number K matrices,arising as representations in some Hilbert spaces,
which may be interpreted as impurity Hilbert spaces.

We now solve (\ref{reflection1}) and (\ref{reflection2}) 
for $K_+(u)$ and $K_-(u)$. For the quantum R matrix (\ref {r}),
One may 
check that the matrix $K_-(u)$ given by
\beq
K_-(u)= \left ( \begin {array}{cccc}
1&0&0&0\\
0&1&0&0\\
0&0&A_-(u)&B_-(u)\\
0&0&C_-(u)&D_-(u)
\end {array} \right ),\label{k-}
\eeq
where
\bea
A_-(u)&=&-(u^2+2u-4c^2_a-8c_a+4u{\bf S}^z_a)/Z_-(u),\no\\
B_-(u)&=&-4u{\bf S}^-_a/Z_-(u),~~~~~~~~ C_-(u)=-4u{\bf S}^+_a/Z_-(u),\no\\
D_-(u)&=&-(u^2+2u-4c^2_a-8c_a-4u{\bf S}^z_a)/Z_-(u),~~~~~~Z_-(u)=(u-2c_a)(u-2c_a-4).
\eea
satisfies (\ref{reflection1}). Here ${\bf S}^{\pm}={\bf S}^x \pm
i{\bf S}^y$.
The matrix $K_+(u)$ can be obtained from the isomorphism of the
superalgebras  ${\cal T}_-  $ and ${\cal T}_+  $. Indeed, given a solution
${\cal T}_- $ of (\ref{reflection1}), then ${\cal T}_+(u)$ defined by
\beq
{\cal T}_+^{st}(u) =  {\cal T}_-(-u)\label{t+t-}
\eeq
is a solution of (\ref{reflection2}). 
The proof follows from some algebraic computations upon
substituting (\ref{t+t-}) into  
(\ref{reflection2}) and making use
of the properties of the R-matrix .
Therefore, one may choose the boundary matrix $K_+(u)$ as 
\beq
K_+(u)=\left ( \begin {array} {cccc}
1&0&0&0\\
0&1&0&0\\
0&0& A_+(u)&B_+(u)\\
0&0&C_+(u)& D_+(u)
\end {array} \right )\label{k^I+}
\eeq
with
\bea
A_+(u)&=&-(u^2-2u-4c^2_b+4+4u{\bf S}^z_b)/Z_+(u),\no\\
B_+(u)&=&-4u{\bf S}^-_b/Z_+(u),~~~~~~~
C_+(u)=-4u{\bf S}^+_b/Z_+(u),\no\\
D_+(u)&=&-(u^2-2u-4c^2_b+4-4u{\bf S}^z_b)/Z_+(u),~~~~~~~Z_+(u)=(u-2c_b+2)(u-2c_b-2).
\eea

Now it can be shown  that 
Hamiltonian (\ref{ham}) is related to the second derivative of the
boundary transfer matrix
$\tau (u)$ with respect to the spectral parameter $u$ at $u=0$ (up to an unimportant additive constant)
\bea
H&=&\frac {t'' (0)}{4(V+2W)}=
  \sum _{j=1}^{L-1} H_{j,j+1} + \frac {1}{2} \stackrel {1}{K'}_-(0)
+\frac {1}{2(V+2W)}\lt[str_0\lt(\stackrel {0}{K}_+(0)G_{L0}\rt)\rt.\no\\
& &\lt.+2\,str_0\lt(\stackrel {0}{K'}_+(0)H_{L0}\rt)+
  str_0\lt(\stackrel {0}{K}_+(0)\lt(H_{L0}\rt)^2\rt)\rt],\label{derived-h}
\eea
where  
\bea
V=str_0 K'_+(0),
~~~~~W=str_0 \lt(\stackrel {0}{k}_+(0) H_{L0}^R\rt),~~~~~~~
H_{i,j}=P_{i,j}R'_{i,j}(0),
~~~~~G_{i,j}=P_{i,j}R''_{i,j}(0).
\eea
This implies that the model under study admits
an infinite number
of conserved currents which are in involution with each other, thus
assuring its integrability.

The Bethe ansatz equations 
 may be derived using the algebraic Bethe ansatz method \cite {Skl88,Gon94,ZM98},
\bea
(\frac {u_j- 1}{u_j+1})^{2L}=
\prod ^N_{\stackrel {i=1}{i \neq j}} \frac {(u_j-u_i -2)(u_j+u_i-2)}
{(u_j-u_i +2)(u_j+u_i+2)}&&\prod ^{M_1}_{\a =1} \frac
{(u_j-v_\a+1)
(u_j+v_\a+1)}
{(u_j-v_\a-1)
(u_j+v_\a-1)},\no\\
\prod_{g =a,b}
\frac{c_g +\frac {v_\a}{2}+1}{c_g-\frac{v_\a}{2}+1}
\prod ^{N}_{j=1} \frac {(v_\a -u_j+1)(v_\a+u_j+1)}
{(v_\a -u_j-1)(v_\a+u_j-1)}
& = & \prod_{\g=1}^{M_2} \frac{(v_\a -w_\g+1)(v_\a +w_\g +1)}
{(v_\a -w_\g-1)(v_\a +w_\g -1)},\no\\
\prod_{g =a,b}\frac{c_g-\frac {w_\g}{2}+\frac {1}{2}}
{c_g-\frac {w_\g}{2}-\frac {1}{2}}
\frac{c_g+\frac {w_\g}{2}-\frac {1}{2}}
{c_g+\frac {w_\g}{2}+\frac {1}{2}}
\prod_{\a=1}^{M_1} \frac{(w_\g -v_\a -1)}
{(w_\g -v_\a +1)}
\frac{(w_\g +v_\a -1)}
{(w_\g +v_\a +1)}
   &=&\prod _{\stackrel {\d=1}{\d \neq \g}}^{M_2}
   \frac {(w_\g-w_\d-2)}{(w_\g -w_\d +2)}
   \frac {(w_\g+w_\d-2)}{(w_\g +w_\d +2)}
  ,\label{Bethe-ansatz}
\eea
with the corresponding energy eigenvalue $E$ of the model 
\beq
E=-\sum ^N_{j=1} \frac {4}{u_j^2-1}.
\eeq

In conclusion, we have studied an integrable Kondo problem describing two
impurities coupled to the 1D supersymmetric extended Hubbard open chain. 
The  quantum integrability of the
system follows from the fact
that the Hamiltonian may be embbeded into
a one-parameter family of commuting transfer matrices. Moreover, the Bethe
Ansatz equations are derived by means of the algebraic Bethe ansatz
approach. It should be emphasized that the boundary K matices found here
are highly nontrivial,since they can not be factorized into the product
of a c-number K matrix and the local momodromy matrices. However,it is
still possible to introduce a singular local monodromy matrix $\tilde
L(u)$
and  express the boundary K
matrix $K_-(u)$ as,
\beq
K_-(u)=\tilde {L}(u){\tilde {L}}^{-1}(-u),
\eeq
where
\beq
\tilde L (u) =
 \left ( \begin {array}
{cccc}
\e & 0&0 &0\\
0& \e &0 &0\\
0& 0& u+2c_a+2 +2{\bf S}^z&2{\bf S}^-\\
0& 0&2 {\bf S}^+&u+2c_a+2-2{\bf S}^z\\
\end {array} \right ).\label{tl}
\eeq
which constitutes a realization of the Yang-Baxter algebra (\ref
{rtt-ttr}) when $\e$
tends to $0$.
The implication of such a singular factorization deserves further
investigation.
Indeed,this implies that integrable Kondo impurities discussed here
appear to be ,in some sense,related to a singular realization of
the Yang-Baxter algebra,which in turn  reflects a hidden six-vertex XXX
symmetry in the original quantum R matrix. A similar situation also
occurs in the supersymmetric t-J model \cite {ZM98}. Also,the extension of the above
construction to the case of arbitrary impurity spin is straightforward.
It will be interesting to carry out the calculation of
thermodynamic,equilibrium properties of the model under consideration.
Especially,it is desirable to study the finite-size spectrum,which,
together with the predictions of the boundary conformal field
theory,will allow us to draw various critical properties. The details
is deferred to a future publication.

\vskip.3in
This work is supported by OPRS and UQPRS. 


\end{document}